# Chapter 3
# Normalized Information Distance

Paul M. B. Vitányi, Frank J. Balbach, Rudi L. Cilibrasi, and Ming Li

**Abstract** The normalized information distance is a universal distance measure for objects of all kinds. It is based on Kolmogorov complexity and thus uncomputable, but there are ways to utilize it. First, compression algorithms can be used to approximate the Kolmogorov complexity if the objects have a string representation. Second, for names and abstract concepts, page count statistics from the World Wide Web can be used. These practical realizations of the normalized information distance can then be applied to machine learning tasks, expecially clustering, to perform feature-free and parameter-free data mining. This chapter discusses the theoretical foundations of the normalized information distance and both practical realizations. It presents numerous examples of successful real-world applications based on these distance measures, ranging from bioinformatics to music clustering to machine translation.

## 3.1 Introduction

The typical data mining algorithm uses explicitly given features of the data to assess their similarity and discover patterns among them. It also comes with many parameters for the user to tune to specific needs according to the domain at hand. In this chapter, by contrast, we are discussing algorithms that neither use features of the data nor provide any parameters to be tuned, but that nevertheless often outperform algorithms of the aforementioned kind. In addition, the methods presented here are not just heuristics that happen to work, but they are founded in the mathematical theory of Kolmogorov complexity. The problems discussed in this chapter will mostly, yet not exclusively, be clustering tasks, in which naturally the notion of distance between objects plays a dominant role.

Paul M. B. Vitányi
CWI, Kruislaan 413, 1098 SJ Amsterdam, The Netherlands; e-mail: paulv@cwi.nl

Frank J. Balbach
University of Waterloo, Waterloo, ON, Canada; e-mail: fbalbach@uwaterloo.ca
(supported by a postdoctoral fellowship of the German Academic Exchange Service (DAAD))

Rudi L. Cilibrasi
CWI, Kruislaan 413, 1098 SJ Amsterdam, The Netherlands; e-mail: cilibrar@cilibrar.com

Ming Li
University of Waterloo, Waterloo, ON, Canada; e-mail: mli@uwaterloo.ca

There are good reasons to avoid parameter laden methods. Setting the parameters requires an intimate understanding of the underlying algorithm. Setting them incorrectly can result in missing the right patterns or, perhaps worse, in detecting false ones. Moreover, comparing two parametrized algorithms is difficult because different parameter settings can give a wrong impression that one algorithm is better than another, when in fact one is simply adjusted poorly. Comparisons using the optimal parameter settings for each algorithm are of little help because these settings are hardly ever known in real situations. Lastly, tweaking parameters might tempt users to impose their assumptions and expectations on the algorithm.

There are also good reasons to avoid feature based methods. Determining the relevant features requires domain knowledge, and determining how relevant they are often requires guessing. Implementing the feature extraction in an algorithm can be difficult, error-prone, and is often time consuming. It also limits the applicability of an algorithm to a specific field.

How can an algorithm perform well if it does not extract the important features of the data and does not allow us to tweak its parameters to help it do the right thing? Of course, parameter and feature free algorithms cannot mind read, so if we a priori know the features, how to extract them, and how to combine them into exactly the distance measure we want, we should do just that. For example, if we have a list of cars with their color, motor rating, etc. and want to cluster them by color, we can easily do that in a straightforward way.

Parameter and feature free algorithms are made with a different scenario in mind. In this *exploratory data mining* scenario we are confronted with data whose important features and how to extract them are unknown to us (perhaps there are not even features). We are then striving not for a certain similarity measure, but for *the* similarity measure between the objects. Does such an absolute measure of similarity exist at all? Yes, it does, in theory. It is called the information distance, and the idea behind it is that two objects are similar if there is a simple description of how to transform each one of them into the other one. If, however, all such descriptions are complex, the objects are deemed dissimilar. For example, an image and its negative are very similar because the transformation can be described as "invert every pixel." By contrast, a description of how to transform a blank canvas into da Vinci's *Mona Lisa* would involve the complete, and comparably large, description of that painting.

The latter example already points to some issues one has to take care of, like asymmetry and normalization. Asymmetry refers to the fact that, after all, the inverse transformation of the *Mona Lisa* into a blank canvas can be described rather simply. Normalization refers to the fact that the transformation description size must be seen in relation to the size of the participating objects. Section 3.2 details how these and other issues are dealt with and explains in which sense the resulting *information distance* measure is universal. The formulation of this distance measure will involve the mathematical theory of Kolmogorov complexity, which is generally concerned with shortest effective descriptions.

While the definition of the information distance is rather theoretical and cannot be realized in practice, one can still use its theoretical idea and approximate it with practical methods. Two such approaches are discussed in subsequent sections. They differ in which property of the Kolmogorov complexity they use and to what kind of objects they apply. The first approach, presented in Sect. 3.3, exploits the relation between Kolmogorov complexity and data compression and consequently employs common compression algorithms to measure distances between objects. This method is applicable whenever the data to be clustered are given in a compressible form, for instance, as a text or other literal description.

The second approach, presented in Sect. 3.4, exploits the relation between Kolmogorov complexity and probability. It uses statistics generated by common Web search engines to measure distances between objects. This method is applicable to non-literal objects, names and concepts, whose properties and interrelations are given by common sense and human knowledge.

## 3.2 Normalized Information Distance

Kolmogorov complexity measures the absolute information *content* of individual objects. For the purpose of data mining, especially clustering, we would also like to be able to measure the absolute information *distance* between individual objects. Such a notion should be universal in the sense that it contains all other alternative or intuitive notions of computable distances as special cases. Such a notion should also serve as an absolute measure of the informational, or cognitive, distance between discrete objects $x$ and $y$. Such a notion of universal informational distance between two strings is the minimal quantity of information sufficient to translate between $x$ and $y$, generating either string effectively from the other. As a result of the universality requirement, this information distance is uncomputable. However, the study of the abstract properties of such an absolute information distance leads to formulas and approaches applicable in practice, as we will demonstrate in subsequent sections.

In this section, we first give a brief introduction to the theory of Kolmogorov complexity, providing definitions and fundamental results. We then derive the unnormalized information distance and show its universality with respect to unnormalized distance measures. Finally we discuss the normalized information distance.

### 3.2.1 Kolmogorov Complexity

To provide some formal framework, we have to give a brief introduction to the theory of Kolmogorov complexity (a comprehensive treatment of this theory is [19]). In order to give a mathematically rigorous definition of Kolmogorov complexity and related terms, we need a few notations and definitions beforehand. By $\mathcal{N}$, $\mathcal{Q}$, $\mathcal{R}$, and $\mathcal{R}^+$ we denote the set of all natural, rational, real, and non-negative real numbers, respectively. For the cardinality of a set $S$ we write $|S|$. In the following we will only consider binary strings $x \in \{0,1\}^*$. All other objects that we might consider can be encoded in a natural way as such strings. We write $\varepsilon$ for the empty string, and $\ell(x)$ for the length of string $x$. All binary strings can be totally ordered according to their length and within the same length lexicographically. We implicitly identify every string with its number in this ordering. Using this identity, we have $\ell(x) = \lfloor \log(x+1) \rfloor$.

For any string $x$ we denote by $\bar{x}$ the string $\bar{x} = 1^{\ell(x)}0x$, called the self-delimiting encoding of $x$. The set $\{\bar{x} : x \in \{0,1\}^*\}$ then is a prefix set, that is, no element of it is prefix of another element. Prefix sets have an important property, namely they satisfy the *Kraft inequality*:

**Lemma 3.1.** *Let $S \subset \{0,1\}^*$ be a prefix set. Then*

$$\sum_{x \in S} 2^{-\ell(x)} \leq 1 \,.$$

Partial functions whose domain is a prefix set are called prefix functions. They play a major role in the theory of Kolmogorov complexity.

Using $\bar{x}$ one can define a pairing function $\langle x, y \rangle = \bar{x}y$, which can be extended to $k$ strings: $\langle x_1, \ldots, x_k \rangle = \langle x_1, \langle x_2, \langle \ldots \langle x_{k-1}, x_k \rangle \ldots \rangle \rangle \rangle = \bar{x}_1 \ldots \bar{x}_{k-1} x_k$. Functions with more than one argument can be realized in the usual way via the pairing function as $\varphi(x_1, \ldots, x_k) = \varphi(\langle x_1, \ldots, x_k \rangle)$. We use an effective enumeration $\varphi_1, \varphi_2, \ldots$ of all partial recursive functions with one argument, and also an effective enumeration $\psi_1, \psi_2, \ldots$ of all partial recursive prefix functions.

Now everything is in place to formulate the fundamental definition of Kolmogorov theory. Consider a function $\varphi$ and two strings $p$ and $x$ such that $\varphi(p) = x$. The string $p$ can be interpreted as a description of $x$ by means of the *description language* $\varphi$. Of course, the string $x$ can have many such descriptions, among which the shortest ones are special. The length of a shortest description $p$ is called the complexity of the string $x$ with respect to $\varphi$. A slightly more general version of complexity takes into account an additional input $y$. In this generalization the description of $x$ is conditional to $y$:

**Definition 3.1.** Let $\varphi$ be a partial recursive function. The conditional complexity (with respect to $\varphi$) of $x$ given $y$ is defined by

$$C_\varphi(x|y) = \min\{\ell(p) : \varphi(y, p) = x\},$$

the unconditional complexity of $x$ by $C_\varphi(x) = C_\varphi(x|\varepsilon)$.

Clearly the complexity of a string depends on the choice of $\varphi$. There is, however, a distinguished function $\varphi_0$, which essentially assigns the lowest possible values, over all partial recursive functions $\varphi$.

**Theorem 3.1.** *There is a partial recursive function $\varphi_0$ such that for all partial recursive functions $\varphi$ there is a constant $c$ with*

$$C_{\varphi_0}(x|y) \leq C_\varphi(x|y) + c$$

*for all $x$ and $y$.*

It is sufficient for $\varphi_0$ to satisfy $\varphi_0(y, n, p) = \varphi_n(y, p)$. Intuitively the behavior of $\varphi_0$ is to take one of its arguments, $n$, and simulate the $n$-th partial recursive function on the input comprised of $\varphi_0$'s other two arguments. In other words, $\varphi_0$ is a universal function for our enumeration $(\varphi_i)_{i \geq 1}$. The function $\varphi_0$ is not unique, but every such function defines essentially the same complexity $C_{\varphi_0}$, that is, up to an additive constant (this follows from Theorem 3.1). Instead of $C_{\varphi_0}$ one typically simply writes $C$.

A helpful intuition for understanding $C$ is to regard $C(x|y)$ as the length of a shortest computer program, in any popular language, that outputs $x$ on input $y$.

While simple and elegant, the notion of $C(\cdot)$ has some oddities. For instance, $C(xy)$ is in general not upper-bounded by the sum of $C(x)$ and $C(y)$. This is one of the reasons that in many cases it is beneficial not to consider *all* partial recursive functions, but only the prefix functions. A result very similar to Theorem 3.1 holds for these functions.

**Theorem 3.2.** *There is a partial recursive prefix function $\psi_0$ such that for all partial recursive prefix functions $\psi$ there is a constant $c$ with*

$$C_{\psi_0}(x|y) \leq C_\psi(x|y) + c$$

*for all $x$ and $y$.*

Instead of $C_{\psi_0}(x|y)$ and $C_{\psi_0}(x|\varepsilon)$ it is customary to write $K(x|y)$ and $K(x)$, respectively. Also the expression $K(\langle x, y \rangle)$ is usually written simply as $K(x, y)$. We refer to $K$ as the Kolmogorov complexity. Our intuition about values of $K(x|y)$ is essentially the same as for $C(x|y)$, the length of the shortest program that outputs $x$ on input $y$. But in contrast to $C$, all shortest programs that "occur" in $K$ constitute a prefix set. This implies, among other important things, that two such programs can be concatenated and still recognized as two distinct programs. This in turn allows the construction of a program that simulates two other programs and combines their output, at the same time being only a constant number of bits larger than the concatenation of the original two programs. A consequence of this is that for $K$ the subadditivity holds, that is, $K(xy) \leq K(x) + K(y) + O(1)$. The next theorem summarizes some more properties of $K$.

**Theorem 3.3.** *1. K is not partial recursive.*
*2. $K(x) \leq \ell(x) + 2\log \ell(x) + O(1)$ for all x.*
*3. $K(x,y) \leq K(x) + K(y|x) + O(1)$ for all x, y.*
*4. Up to an additional term of $O(1)$:*

$$K(x,y) = K(x) + K(y|\langle x, K(x)\rangle) = K(y) + K(x|\langle y, K(y)\rangle)$$

*for all x, y.*
*5. Up to an additional term of $O(\log K(xy))$:*

$$K(x,y) = K(x) + K(y|x) = K(y) + K(x|y)$$

*for all x, y.*

Item 1 does not come as a surprise since, intuitively, in order to find $K(x)$ one has to verify that no program under a certain length outputs $x$, a task that conflicts with the undecidability of the halting problem.

Item 2 gives an upper bound of $K(x)$ in terms of the length of $x$. In order to describe $x$ prefix free, one can use an advanced self-delimiting encoding of $x$ namely $\bar{\bar{x}} = \overline{\ell(x)}x$, which has the length $\ell(x) + 2\ell(\ell(x)) + 1$.

Items 3, 4, and 5 elaborate on the subadditivity property. While Item 3 only provides a better *upper bound*, which can be easily understood via the intuition of program lengths, the other two items state *equalities* and require sophisticated proofs. These results go by the name *Symmetry of Information* and will prove useful later in Sect. 3.3.1.

All programs that can be identified as shortest ones by $K$ form a prefix set. It follows by the Kraft inequality that $\sum_x 2^{-K(x)} \leq 1$. This means that the values $2^{-K(x)}$ can be regarded as quantities very similar to probabilities, because they sum up to at most 1. In this assignment of values, less complex objects receive a higher probability than more complex objects. This can be seen as a "smooth" compromise between the contrary views of *Occam's Razor*, which advises to consider the simplest explanation only, and Epicurus's *Principle of Multiple Explanations*, which advises to consider all explanations. Indeed, this *algorithmic probability* $R(x) = 2^{-K(x)}$ is universal in a sense made clear in the remainder of this section.

Since in the realm of probabilities we are dealing with real valued functions, we first need to introduce some notions of computability for them.

**Definition 3.2.** A real valued function $f: \mathcal{N} \to \mathcal{R}$ is called *lower semicomputable* if there is a recursive function $g: \mathcal{N} \times \mathcal{N} \to \mathcal{Q}$ such that for all $x$ the series $(g(x,k))_{k \in \mathcal{N}}$ is nondecreasing and $f(x) = \lim_{k \to \infty} g(x,k)$. The function $f$ is called *upper semicomputable* if $-f$ is lower semicomputable.

A semimeasure assigns a non-negative real number to every string (or, equivalently, natural number). It differs from a probability measure in that the sum of all these values can be less than 1. In the same way as there are conditional probabilities, we can also consider conditional semimeasures.

**Definition 3.3.** A *discrete conditional semimeasure* is a function $P: \mathcal{N} \times \mathcal{N} \to \mathcal{R}^+$ such that for all $y$:

$$\sum_x P(x|y) \leq 1 \ .$$

The large class of *lower semicomputable* semimeasures has a universal element **m** that dominates every other lower semicomputable semimeasure up to a multiplicative constant.

**Theorem 3.4.** *There is a lower semicomputable discrete semimeasure* **m** *such that for all lower semicomputable discrete semimeasures P:*

$$P(x|y) = O(\mathbf{m}(x|y)) \;.$$

This universal semimeasure is intimately related to the Kolmogorov complexity via the *Conditional Coding Theorem* [19]:

**Theorem 3.5.** $-\log \mathbf{m}(x|y) = K(x|y) + O(1)$.

### 3.2.2 Information Distance

Intuitively, the minimal information distance between $x$ and $y$ is the length of the shortest program for a universal computer to transform $x$ into $y$ and $y$ into $x$. This program then functions in a "catalytic" manner, being retained in the computer before, during, and after the computation. This measure will be shown to be, up to a logarithmic additive term, equal to the *maximum* of the conditional Kolmogorov complexities. The conditional complexity $K(y|x)$ itself is unsuitable as optimal information distance because it is asymmetric: $K(\varepsilon|x)$ is small for all $x$, yet intuitively a long random string $x$ is not close to the empty string. The asymmetry of the conditional complexity $K(x|y)$ can be remedied by defining the algorithmic informational distance between $x$ and $y$ to be the sum of the relative complexities, $K(y|x) + K(x|y)$. The resulting metric will overestimate the information required to translate between $x$ and $y$ in case there is some redundancy between the information required to get from $x$ to $y$ and the information required to get from $y$ to $x$.

For a partial recursive function $\varphi$, let

$$E_\varphi(x,y) = \min\{\ell(p) : \varphi(p,x) = y \text{ and } \varphi(p,y) = x\} \;.$$

**Lemma 3.2.** *There is a universal partial recursive prefix function $\psi_0$ such that for each partial recursive prefix function $\psi$ and all $x,y$,*

$$E_{\psi_0}(x,y) \leq E_\psi(x,y) + c_\psi \;,$$

*where $c_\psi$ is a constant that depends on $\psi$ but not on $x$ and $y$.*

By Lemma 3.2, for every two universal prefix functions $\varphi_0$ and $\psi_0$, we have for all $x,y$ that $|E_{\varphi_0}(x,y) - E_{\psi_0}(x,y)| \leq c$, with $c$ a constant depending on $\varphi_0$ and $\psi_0$ but not on $x$ and $y$. Thus the following definition is machine-independent.

**Definition 3.4.** Fixing a particular universal prefix function $\psi_0$, we define *information distance* as

$$E_0(x,y) = \min\{\ell(p) : \psi_0(p,x) = y \text{ and } \psi_0(p,y) = x\} \;. \tag{3.1}$$

**Definition 3.5.** The *max distance* between $x$ and $y$ is $E(x,y) = \max\{K(x|y), K(y|x)\}$.

It has been proved that up to an additive logarithmic term, the information distance $E_0$ is equal to the max distance.

#### 3.2.2.1 Maximal Overlap

To what extent can the information required to compute $x$ from $y$ be made to overlap with that required to compute $y$ from $x$? In some simple cases, complete overlap can be achieved, so that the same minimal program suffices to compute $x$ from $y$ as to compute $y$ from $x$.

*Example 3.1.* If $x$ and $y$ are independent random binary strings of the same length $n$ (up to additive constants $K(x|y) = K(y|x) = n$), then their bitwise exclusive-or $x \oplus y$ serves as a minimal program for both computations. Similarly, if $x = uv$ and $y = vw$ where $u$, $v$, and $w$ are independent random strings of the same length, then $u \oplus w$ is a minimal program to compute either string from the other.

Now suppose that more information is required for one of these computations than for the other, say, $K(y|x) > K(x|y)$. Then the minimal programs cannot be made identical because they must be of different sizes. Nevertheless, in simple cases, the overlap can still be made complete, in the sense that the larger program (for $y$ given $x$) can be made to contain all the information in the smaller program, as well as some additional information. This is so when $x$ and $y$ are independent random strings of unequal length, for example $u$ and $vw$ above. Then $u \oplus v$ serves as a minimal program for $u$ from $vw$, and $(u \oplus v)w$ serves as one for $vw$ from $u$.

The following *Conversion Theorem* asserts the existence of a difference string $p$ of length $\ell(p) = \max\{K(x|y), K(y|x)\}$, up to an additive logarithmic term, that converts both ways between $x$ and $y$ and at least one of these conversions is optimal. If $K(x|y) = K(y|x)$, then the conversion is optimal in both directions.

**Theorem 3.6.** *Let $x$ and $y$ be strings such that $K(y|x) \geq K(x|y)$. There is a string $r$ of length $K(y|x) - K(x|y)$ such that*

$$E_0(rx, y) = K(x|y) + K(K(x|y), K(y|x)) + O(1) \ .$$

**Corollary 3.1.** $E_0(x, y) = \max\{K(x|y), K(y|x)\} + O(\log \max\{K(x|y), K(y|x)\})$.

#### 3.2.2.2 Universality

Let us assume we want to quantify how much some given objects differ with respect to a specific feature, for instance, the length of files in bits, the number of beats per second in music pieces, or the number of occurrences of some base in genomes. Every specific feature induces a specific distance measure, and conversely every distance measure can be viewed as a quantification of a feature difference.

Every distance measure should be an effectively approximable positive function of the two objects that satisfies a reasonable density condition and obeys the triangle inequality. It turns out that $E$ is minimal up to an additive constant among all such distances. Hence, it is a universal *information distance* that accounts for any effective resemblance between two objects.

Let us consider an example of measuring *distance* between two pictures. Identify digitized black-and-white pictures with binary strings. There are many distances defined for binary strings, for example, the Hamming distance and the Euclidean distance. Such distances are sometimes appropriate. For instance, if we take a binary picture and change a few bits on that picture, then the changed and unchanged pictures have small Hamming or Euclidean distance, and they do look similar.

However, these measures are not always appropriate. The positive and negative prints of a photo have the largest possible Hamming and Euclidean distance, yet they look similar. Also, if we shift a picture one bit to the right, again the Hamming distance may increase by a lot, but the two pictures remain similar.

Many approaches to pattern recognition define distance measures with respect to pictures, language sentences, vocal utterances, and many more. We have already seen evidence that $E(x,y) = \max\{K(x|y), K(y|x)\}$ is a natural way to formalize a notion of algorithmic informational distance between $x$ and $y$. Let us now show that the distance $E$ is, in a sense, minimal among all reasonable distance measures.

In general we differentiate between distance functions and metrics. The latter are distance measures that satisfy additional conditions as formalized in the following.

**Definition 3.6.** A *distance function* is a function $D\colon \{0,1\}^* \times \{0,1\}^* \to \mathscr{R}^+$. It is a *metric* if it satisfies the metric (in)equalities:

- $D(x,y) = 0$ if and only if $x = y$, (identity)
- $D(x,y) = D(y,x)$, (symmetry)
- $D(x,y) \leq D(x,z) + D(z,y)$. (triangle inequality)

The value $D(x,y)$ is called the *distance* between $x$ and $y$. As a familiar example of a distance function that is also a metric, consider the Euclidean metric, the everyday distance $D_E(a,b)$ between two geographical objects $a,b$ expressed in, say, meters. Clearly, this distance satisfies the properties $D_E(a,a) = 0$, $D_E(a,b) = D_E(b,a)$, and $D_E(a,b) \leq D_E(a,c) + D_E(c,b)$ (for instance, $a$ = Amsterdam, $b$ = Beijing, and $c$ = Chicago.) Our goal is to generalize this concept of distance from our physical space to the cyberspace and characterize the set of all reasonable distance functions that would measure informational distances between objects.

For a distance function or metric to be reasonable, it has to satisfy a certain additional condition, referred to as *density condition*. Intuitively this means that for every object $x$ and value $d \in \mathscr{R}^+$ there is at most a certain, finite number of objects $y$ at distance $d$ from $x$. This requirement excludes degenerate distance measures like $D(x,y) = 1$ for all $x \neq y$. Exactly how fast we want the distances of the strings $y$ from $x$ to go to infinity is not important, it is only a matter of scaling. For convenience, we will require the following *density conditions*:

$$\sum_{y:y\neq x} 2^{-D(x,y)} \leq 1 \qquad \text{and} \qquad \sum_{x:x\neq y} 2^{-D(x,y)} \leq 1 \,. \tag{3.2}$$

Finally we allow only distance measures that are computable in some broad sense, which will not be seen as unduly restrictive. More precisely, only upper semicomputability of $D$ will be required. This is reasonable: as we have more and more time to process $x$ and $y$ we may discover newer and newer similarities among them, and thus may revise our upper bound on their distance. The next definition summarizes the class of distance measures we are concerned with.

**Definition 3.7.** An *admissible information distance* is a total, possibly asymmetric, nonnegative function on the pairs $x,y$ of binary strings that is 0 if and only if $x = y$, is upper semicomputable, and satisfies the density requirement (3.2).

*Example 3.2.* The Hamming distance between two strings $x = x_1\ldots x_n$ and $y = y_1\ldots y_n$ is defined as $d(x,y) = |\{i : x_i \neq y_i\}|$. This distance does not directly satisfy the density requirements (3.2). With minor modification, we can scale it to satisfy these requirements. In representing the Hamming distance $m$ between $x$ and $y$, strings of equal length $n$ differing in positions $i_1,\ldots,i_m$, we can use a simple prefix-free encoding of $(n,m,i_1,\ldots,i_m)$ in $H_n(x,y) = 2\log n + 4\log\log n + 2 + m\log n$ bits. We encode $n$ and $m$ prefix-free in $\log n + 2\log\log n + 1$ bits each and then the literal indexes of the actual flipped-bit positions. Thus, $H_n(x,y)$ is the length of a prefix code word specifying the positions where $x$ and $y$ differ. This modified Hamming distance is symmetric, and it is an admissible distance by the Kraft inequality $\sum_{y:y\neq x} 2^{-H_n(x,y)} \leq 1$. It is easy to verify that $H_n$ is a metric in the sense that it satisfies the metric (in)equalities up to $O(\log n)$ additive precision.

The following theorem is the fundamental result about the max distance $E$. It states that $E$ is an optimal admissible information distance.

**Theorem 3.7.** *The function $E$ with $E(x,y) = \max\{K(x|y), K(y|x)\}$ is an admissible information distance and a metric. It is minimal in the sense that for every admissible information distance $D$, we have $E(x,y) \leq D(x,y) + O(1)$.*

The quantitative difference in a certain feature between two objects can be considered as an admissible distance. Theorem 3.7 shows that the information distance $E$ is universal in that among all admissible distances it is always least. That is, it accounts for the dominant feature in which two objects are alike. For that reason $E$ is also called the *universal information distance*.

Many admissible distances are absolute, but if we want to express similarity, then we are more interested in relative ones. For example, if two strings of $1,000,000$ bits differ by $1,000$ bits, then we are inclined to think that those strings are relatively similar. But if two strings of $1,000$ bits differ by $1,000$ bits, then we find them very different.

*Example 3.3.* Consider the problem of comparing genomes. The *E. coli* genome is about 4.8 megabase long, whereas *H. influenza*, a sister species of *E. coli*, has genome length only 1.8 megabase. The information distance $E$ between the two genomes is dominated by their length difference rather than the amount of information they share. Such a measure will trivially classify *H. influenza* as being closer to a more remote species of similar genome length such as *A. fulgidus* (2.18 megabase), rather than with *E. coli*. In order to deal with such problems, we need to normalize.

Our objective now is to normalize the universal information distance $E(x,y) = \max\{K(x|y), K(y|x)\}$ to obtain a universal *similarity* distance. It should give a similarity with distance 0 when objects are maximally similar and distance 1 when they are maximally dissimilar.

### 3.2.3 Normalized Information Distance

It is paramount that the normalized version of the universal information distance metric is also a metric. Were it not, then the relative relations between the objects in the space would be disrupted and this could lead to anomalies, if, for instance, the triangle inequality would be violated for the normalized version.

In order to obtain a normalized universal information distance function, both versions of information distance discussed so far, $E_0$ and $E$, can be normalized. We will only discuss how to normalize the max distance $E$ and call it the *normalized information distance*.

**Definition 3.8.** The *normalized information distance* (NID) between two strings $x$ and $y$ is defined as

$$e(x,y) = \frac{\max\{K(x|y), K(y|x)\}}{\max\{K(x), K(y)\}} \ . \tag{3.3}$$

Dividing by $\max\{K(x), K(y)\}$ is not the most obvious idea for normalizing $E$, but the more obvious ideas do not work:

- Divide by the length. Then no matter whether we divide by the sum or maximum of the length, the triangle inequality is not satisfied.

- Divide by $K(x,y)$. Then the distances will be $1/2$ whenever $x$ and $y$ satisfy $K(x) \approx K(y) \approx K(x|y) \approx K(y|x)$. In this situation, however, $x$ and $y$ are completely dissimilar, and we would expect distance values of about 1.

That the NID is indeed a normalized metric is a remarkable fact [18].

**Theorem 3.8.** *The normalized information distance $e(x,y)$ takes values in the range $[0,1]$ and is a metric, up to ignorable discrepancies.*

This concludes our discussion of the theoretical foundations of the NID. We continue with demonstrations of how these insights can be put to use in practical settings.

### 3.3 Normalized Compression Distance

The normalized information distance is theoretically appealing, but impractical since it cannot be computed. In this section we discuss the normalized *compression* distance, an efficiently computable, and thus practically applicable, form of the normalized information distance.

#### 3.3.1 Introduction

The normalized information distance $e(x,y)$, is called *universal* because it accounts for the dominant difference between two objects. It depends on the uncomputable function $K$, and is therefore also uncomputable. First we observe that using $K(x,y) = K(xy) + O(\log \min\{K(x), K(y)\})$ and Item 5 of Theorem 3.3 we obtain

$$E(x,y) = \max\{K(x|y), K(y|x)\} = K(xy) - \min\{K(x), K(y)\}, \tag{3.4}$$

up to an additive logarithmic term $O(\log K(xy))$ which we ignore in the sequel.

By rewriting $E$ as in (3.4) we manage to remove all conditional complexity terms and obtain a formula with only the non-conditional terms $K(x), K(y), K(xy)$. This comes in handy if we interpret $K(x)$ as the length of the string $x$ after being maximally compressed. With this in mind, it is an obvious idea to approximate $K(x)$ with the length of the string $x$ under an efficient real-world compressor. Any correct and lossless data compression program can provide an upper-bound approximation to $K(x)$, and most good compressors detect a large number of statistical regularities.

Substituting the numerator of (3.3) with (3.4) and subsequently using a real-world compressor $Z$ (such as `gzip`, `bzip2`, `PPMZ`) to heuristically replace the Kolmogorov complexity, we obtain the distance $e_Z$, often called the *normalized compression distance* (NCD), defined by

$$e_Z(x,y) = \frac{Z(xy) - \min\{Z(x), Z(y)\}}{\max\{Z(x), Z(y)\}}, \tag{3.5}$$

where $Z(x)$ denotes the binary length of the compressed version of the string $x$ compressed with compressor $Z$. The distance $e_Z$ is actually a family of distances parametrized with the compressor $Z$. The better $Z$ is, the closer $e_Z$ approaches the normalized information distance, the better the results are expected to be.

Under mild conditions on compressor $Z$, the distance $e_Z$ is computable, takes values in $[0,1]$, and is a metric [9]. More formally, a compressor $Z$ is *normal* if it satisfies the axioms

- $Z(xx) = Z(x)$ and $Z(\varepsilon) = 0$, (identity)
- $Z(xy) \geq Z(x)$, (monotonicity)
- $Z(xy) = Z(yx)$, (symmetry)
- $Z(xy) + Z(z) \leq Z(xz) + Z(yz)$, (distributivity)

up to an additive $O(\log n)$ term, with $n$ the maximal binary length of a string involved in the (in)equality concerned.

Then the unnormalized distance $E_Z(x,y) = Z(xy) - \min\{Z(x), Z(y)\}$, with $Z$ a normal compressor, is computable, satisfies the density requirement in (3.2), and satisfies the metric (in)equalities up to additive $O(\log n)$ terms, with $n$ the maximal binary length of a string involved in the (in)equality concerned.

Moreover, the normalized distance $e_Z$ of (3.5), with $Z$ a normal compressor, has values in $[0,1]$ and satisfies the metric (in)equalities up to additive $O((\log n)/n)$ terms, with $n$ the maximal binary length of a string involved in the (in)equality concerned.

Informal experiments [9] have shown that these axioms are in various degrees satisfied by good real-world compressors like **bzip2**, and **PPMZ**, with **PPMZ** being best among the ones tested. The compressor **gzip** performed not so well, and in all cases some compressor-specific window or block size determines the maximum useable length of the arguments $x$ and $y$ (32 KB for **gzip**, 450 KB for **bzip2**, unlimited for **PPMZ**). Cebrián et al. [4] systematically investigated how far the performance of real-world compressors **gzip**, **bzip2**, and **PPMZ** satisfy the identity axiom $Z(xx) = Z(x)$ of a normal compressor.

The normalized information distance $e$ is intended to be universally applicable. In practice, various computable distances, including $e_Z$, can be viewed as approximations to $e$. Moreover, many of the measures used in the data mining community (see Tan et al. [24]) may, after normalization, be viewed as various degrees of approximations to $e$.

The NCD has been put to numerous tests. Keogh et al. [14; 15] have tested a closely related metric as a parameter-free and feature-free data mining tool on a large variety of sequence benchmarks. Comparing the NCD method with 51 major parameter-loaded methods found in the eight major data-mining conferences (SIGKDD, SIGMOD, ICDM, ICDE, SSDB, VLDB, PKDD, and PAKDD) in the last decade, on all data bases of time sequences used, ranging from heart beat signals to stock market curves, they established clear superiority of the NCD method for clustering heterogeneous data, and for anomaly detection, and competitiveness in clustering domain data.

Apart from providing a theoretical justification for these practical distances, the normalized information distance does more in that it embodies all approximations. The broad range of successful applications of $e_Z$ will be demonstrated in the remainder of this section, where we will discuss applications in bioinformatics, linguistics, music, and plagiarism detection.

### 3.3.2 Phylogenies

DNA sequences seem ideally suited for the compression distance approach. A DNA sequence is a finite string over a four-letter alphabet $\{A, C, G, T\}$. We used the whole mitochondrial genomes of 20 mammals, each of about 18,000 base pairs, to test a hypothesis of Eutherian orders. It has been hotly debated in biology which two of the three main placental mammalian groups, Primates, Ferungulates, and Rodents, are more closely related. One cause of the debate is that the standard maximum likelihood method, which depends on the multiple alignment of sequences corresponding to an individual protein, gives (Rodents, (Ferungulates,

Primates)) for half of the proteins in the mitochondrial genome, and (Ferungulates, (Primates, Rodents)) for the other half.

In recent years, when people use more sophisticated methods, together with biological evidences, it is believed that (Rodents, (Ferungulates, Primates)) reflects the true evolutionary history. We confirm this from the whole genome perspective using the distance $e_Z$. We use the complete mitochondrial genome sequences from following 20 species: rat (*Rattus norvegicus*), house mouse (*Mus musculus*), gray seal (*Halichoerus grypus*), harbor seal (*Phoca vitulina*), cat (*Felis catus*), white rhino (*Ceratotherium simum*), horse (*Equus caballus*), finback whale (*Balaenoptera physalus*), blue whale (*Balaenoptera musculus*), cow (*Bos taurus*), gibbon (*Hylobates lar*), gorilla (*Gorilla gorilla*), human (*Homo sapiens*), chimpanzee (*Pan troglodytes*), pygmy chimpanzee (*Pan paniscus*), orangutan (*Pongo pygmaeus*), Sumatran orangutan (*Pongo pygmaeus abelii*), with opossum (*Didelphis virginiana*), wallaroo (*Macropus robustus*) and platypus (*Ornithorhynchus anatinus*) as the outgroup.

For every pair of mitochondrial genome sequences *x* and *y*, evaluate the formula in (3.5) using a special-purpose DNA sequence compressor **DNACompress** [7], or a good general-purpose compressor like **PPMZ**. The resulting distances are the entries in a $20 \times 20$ distance matrix. Constructing a phylogeny tree from the distance matrix, using common tree reconstruction software, gives the tree in Fig. 3.1. This tree confirms the accepted hypothesis of (Rodents, (Primates, Ferungulates)), and every single branch of the tree agrees with the current biological classification.

Similarity of sequences in biology is currently primarily handled using alignments. However, the alignment methods seem inadequate for post-genomic studies since they do not scale well with data set size and they seem to be confined only to genomic and proteomic sequences. Therefore, alignment-free similarity measures are actively pursued. Ferragina et al. [13] experimentally tested the normalized information distance using 25 compressors to obtain the NCD, and six data sets of relevance to molecular biology. They compared the methodology with methods based on alignments and not. They assessed the intrinsic ability of the methodology to discriminate and classify biological sequences and structures. The compression program **PPMd**, based on **PPM** (Prediction by Partial Matching), for generic data and **GenCompress** [17] for DNA, are the best performers among the compression algorithms they used. The quantitative analysis supports the conclusion that the normalized information / compression method is worth using because of its robustness, flexibility, scalability, and competitiveness with existing techniques. In particular, the methodology applies to all biological data in textual format.

### 3.3.3 Language Trees

The similarity between languages can, to some extent, be determined by the similarity of their vocabulary. This means that given two translations of the same text in different languages, one can estimate the similarity of the languages by the similarity of the words occurring in the translations. This has been exploited by Benedetto et al. [2], who use a compression method related to NCD to construct a language tree of 52 Euroasian languages from translations of the Universal Declaration of Human Rights [1].

In this section we present an experiment [9] that uses the NCD method with translations of the Universal Declaration of Human Rights into 16 languages. Among these languages are four European (German, English, Spanish, Dutch), eight African (Pemba, Dendi, Ndebele, Kicongo, Somali, Rundi, Ditammari, Dagaare), and four American (Chikasaw, Purhepecha, Mazahua, Zapoteco).

The files have been left in their original UTF-8 encoding, and all pairwise distances between them have been determined using $e_Z$, where $Z$ has been chosen to be the standard text compressor **gzip**. From the resulting matrix of distances, the tree in Fig. 3.2 has been generated. It shows the three main language groups, only Dendi and Somali are somewhat too close to the American languages. Also, the classification of English as a Romance language is erroneous from a historic perspective and is due to the English vocabulary being heavily influenced by French and Latin. Therefore the vocabulary, on which the approach discussed here is based, is indeed to a large part Romance.

**Fig. 3.1** The evolutionary tree built from complete mitochondrial DNA sequences of several mammals

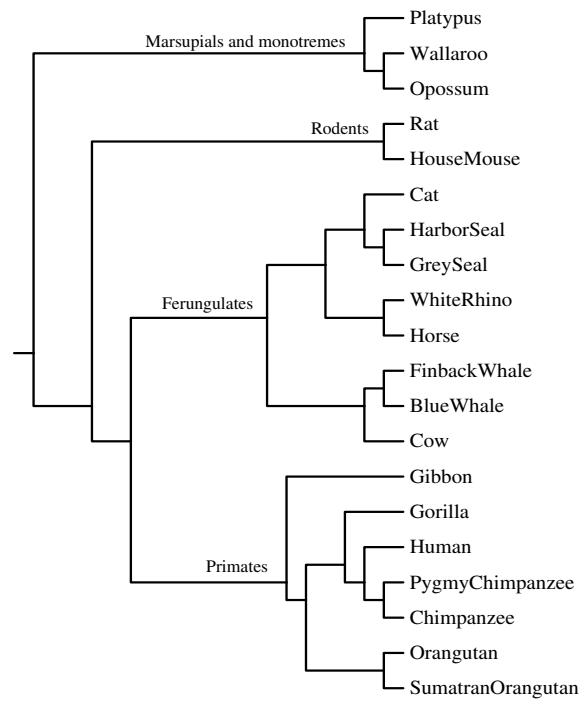

Similar experiments have been conducted with other clustering methods or other languages [18; 9], but with equally plausible results.

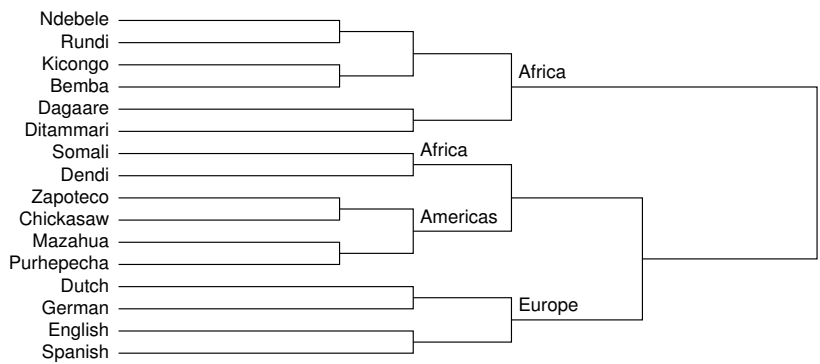

**Fig. 3.2** Language tree based on the Universal Declaration of Human Rights constructed using the NCD, based on the `gzip` compressor

### 3.3.4 Plagiarism Detection

It is a common observation in university courses with programming assignments that some programs are plagiarized from others. That means that large portions are copied from other programs. What makes this hard to detect is that it is relatively easy to change a program syntactically without changing its semantics, for example, by renaming variables and functions, inserting dummy statements or comments, or reordering obviously independent statements. Nevertheless a plagiarized program is somehow close to its source and therefore the idea of using a distance measure on programs in order to uncover plagiarism is obvious.

We briefly describe the SID system [6] that uses a variant of the NID to measure the distance between two source code files. This variant is called the *sum distance*, and its Kolmogorov theoretic formulation is

$$e_{\text{sum}}(x,y) = \frac{K(x|y) + K(y|x)}{K(x,y)} \ .$$

This function takes values in the interval $[0,1]$ and is a metric according to Definition 3.6 [17].

To compute the similarity between two Java source code files, SID first tokenizes the programs and then uses a customized compressor to approximate the sum distance. This compressor is a variant of the Lempel-Ziv compression scheme [26] that has an unbounded buffer size and can thus detect repetitions over the entire file. Moreover it also takes advantage of *approximate* repetitions to increase the compression rate.

Evaluating plagiarism detection systems is difficult, but field experiments indicate that SID performs competitively and is more robust against certain attempts to circumvent detection, such as insertion of irrelevant code.

### 3.3.5 Clustering Music

The previous examples of NCD applications were based on text, be it source code or the Declaration of Human Rights. But The NCD method can also be applied to multimedia data like music, if it is present in the right format.

Music files in the MIDI format can be transformed into files that can be successfully clustered with the NCD. This transformation involves stripping the files of all instrument indicators, MIDI control signals and meta information such as title and composer. What essentially remains of a file is a list of musical notes of the piece. These preprocessed files can than be treated as text files.

A number of experiments has been performed [9; 10] with such files. We present a single, representative one, in which the set of musical pieces comprises four preludes from Chopin's Opus 28, two preludes and two fugues from Bach's "Das wohltemperierte Klavier," and the four movements from Debussy's "Suite Bergamesque." After preprocessing the MIDI files as described above, the pairwise $e_Z$ values, with `bzip2` as compressor, are computed. To generate the final hierarchical clustering as shown in Fig. 3.3, a special quartet method [9; 10] is used.

Perhaps with the exception of Chopin's Prélude no. 5, which seems somewhat closer to the Bach pieces, the results agree with one's expectations.

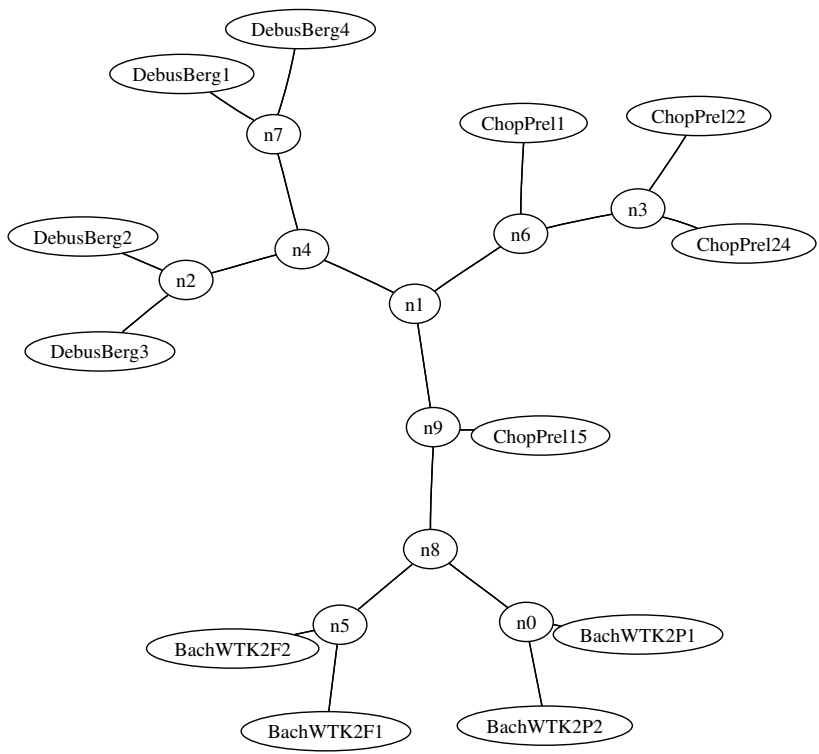

**Fig. 3.3** Hierarchical clustering of MIDI files of four pieces by Bach, Chopin, and Debussy using the `bzip2` based NCD and a quartet method

### 3.3.6 Clustering Heterogeneous Data

We test gross classification of files based on heterogeneous data of markedly different file types:

1. Four mitochondrial gene sequences, from a black bear, polar bear, fox, and rat obtained from the GenBank Database [3] on the World Wide Web.
2. Four excerpts from the novel *The Zeppelin's Passenger* by E. Phillips Oppenheim, obtained from the Project Gutenberg Edition on the World Wide Web.
3. Four MIDI files without further processing: two from Jimi Hendrix and two movements from Debussy's "Suite Bergamasque," downloaded from various repositories on the World Wide Web.
4. Two Linux x86 ELF executables (the `cp` and `rm` commands), copied directly from the RedHat 9.0 Linux distribution.
5. Two compiled Java class files, generated by ourselves.

The program correctly classifies each of the different types of files together with like near like. The result is reported in Fig. 3.4. This experiment shows the power and universality of the method: no features of any specific domain of application are used. We believe that there is no other method known that can cluster data that are so heterogeneous this reliably.

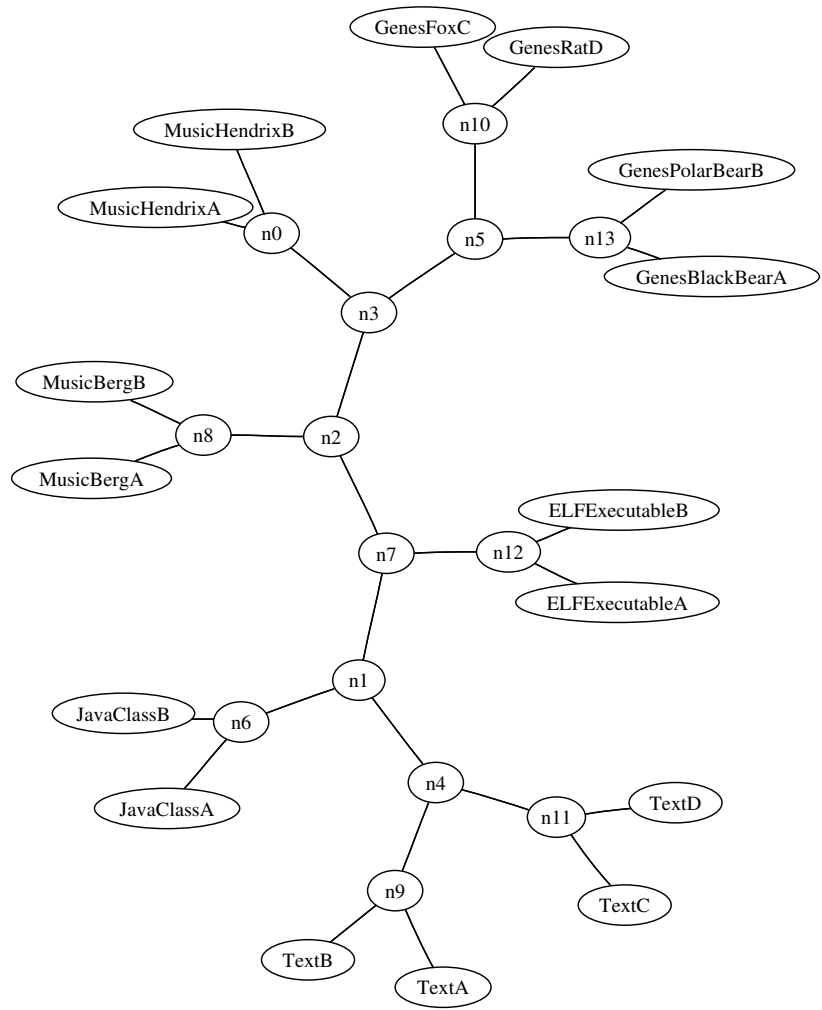

**Fig. 3.4** Clustering of heterogeneous file types using the NCD, based on the `bzip2` compressor, and a quartet clustering method. The set of file contains four MIDI files, four genomes, four English texts, and two Java class files and Linux executables

### 3.3.7 Conclusion

The NCD is universal, in a mathematical sense as approximation of the universal NID, but also in a practical sense, as witnessed by the wide range of successful applications. Nevertheless the practical universality is of a different flavor because the NCD is a family of distance measures parametrized by a compressor. This means that one has to pick a suitable compressor for the application domain at hand. It does, however, not mean that one has to know the relevant features of the objects in that domain beforehand. Rather, using a good compressor for objects in a certain domain, makes it more likely that the compressor does indeed capture the (combinations of) features relevant for the objects in that domain. The criterion for choosing a compressor is clearer and simpler than the criterion for picking the "right" features, namely encoding length.

In other words, the NCD is sensitive to many features and discovers the ones that are important for the objects under consideration. Not being tuned to specific features beforehand contributes to the robustness of the NCD method, as well as to the ease of use. It is thus a valuable tool in the process of *exploratory data mining* [14].

## 3.4 Normalized Web Distance

The normalized compression distance can only be applied to objects that are strings or that at least can be naturally represented as such. Abstract concepts or ideas, on the other hand, are not amenable to the NCD method. In this section, we present a realization of NID overcoming that limitation by taking advantage of the World Wide Web.

### 3.4.1 Introduction

There are literal objects and non-literal objects. Examples of the former include the four-letter human genome, the text of *War and Peace* by Tolstoy, and the source code of a program. Non-literal objects are essentially names, either for literal objects, like "the four-letter human genome," "the text of *War and Peace* by Tolstoy," and "`main.c`," or for concepts and ideas that are not associated with a literal object in any way, like the concept of "home" or "red." The latter objects acquire their meaning from their contexts in background common knowledge in humankind. Put differently, a sequence contains information within itself, whereas names and concepts contain their information not within themselves. The name "human genome" implies three gigabases of information. The phrase "*War and Peace* by Tolstoy" perhaps carries information even beyond the book.

Let $\mathbf{W}$ be the set of pages of the World Wide Web, and let $\mathbf{x} \subseteq \mathbf{W}$ be the set of pages containing the search term $x$. By the Conditional Coding Theorem we have $\log 1/\mathbf{m}(\mathbf{x}|\mathbf{x} \subseteq \mathbf{W}) = K(\mathbf{x}|\mathbf{x} \subseteq \mathbf{W}) + O(1)$, where $\mathbf{m}$ is the universal lower semicomputable discrete semimeasure. This equality relates the incompressibility of the set of pages on the Web, containing a given search term, to its universal probability. While we do not know how to compute $\mathbf{m}$, a natural heuristic now is to use the distribution of $x$ in the Web to approximate $\mathbf{m}(\mathbf{x}|\mathbf{x} \subseteq \mathbf{W})$. (We give a simplified approach where we assume that every page contains at most one search term.) Let us define the probability mass function $g(x)$ to be the probability that the search term $x$ appears in a page indexed by a given internet search engine $G$, that is, the number of pages returned divided by the overall number of pages indexed. Then the Shannon-Fano code [23] associated with $g$ can be set at

$$G(x) = \log \frac{1}{g(x)} .$$

Replacing $Z(x)$ by $G(x)$ in the formula in (3.5), we obtain the distance $e_G$, often called the *normalized Web distance* (NWD):

$$e_G(x,y) = \frac{G(xy) - \min\{G(x), G(y)\}}{\max\{G(x), G(y)\}} \tag{3.6}$$

$$= \frac{\max\{\log f(x), \log f(y)\} - \log f(x,y)}{\log N - \min\{\log f(x), \log f(y)\}}.$$

where $f(x)$ is the number of pages containing $x$, the frequency $f(x,y)$ is the number of pages containing both $x$ and $y$, and $N$ is the total number of indexed pages. We can view the search engine $G$ as a compressor using the Web, and $G(x)$ as the binary length of the compressed version of the set of all pages containing the search term $x$, given the indexed pages on the Web. The distance $e_G$ is actually a family of distances parametrized with the search engine $G$. It was originally called normalized Google distance (NGD) and thus featured a particular search engine [11]. The name *normalized Web distance* is more generic and more in line with the name NCD, which also does not mention a concrete compressor.

*Example 3.4.* We describe an experiment, using a popular search engine, performed in the year 2004, at which time it indexed $N = 8,058,044,651$ pages. A search for "horse" returns a page count of 46,700,000. A search for "rider" returns a page count of 12,200,000. A search for both "horse" and "rider" returns a page count of 2,630,000. Thus $e_G(\text{horse}, \text{rider}) = 0.443$. It is interesting to note that this number stayed relatively fixed as the number of pages indexed by the used search engine increased.

The distance $e_G$ is actually a family of distances parametrized with the search engine $G$. The better $G$ is, the closer the $e_G$ approaches the normalized information distance, and the better the results are expected to be. The distance $e_G$ is computable, takes values primarily (but not exclusively) in $[0,1]$, and is symmetric, that is, $e_G(x,y) = e_G(y,x)$. It only satisfies "half" of the identity property, namely $e_G(x,x) = 0$ for all $x$, but $e_G(x,y) = 0$ can hold even if $x \neq y$, for example, if the terms $x$ and $y$ always occur together.

The NWD also does *not* satisfy the triangle inequality $e_G(x,y) \leq e_G(x,z) + e_G(z,y)$ for all $x,y,z$. To see that, choose $x$, $y$, and $z$ such that $x$ and $y$ never occur together, $z$ occurs exactly on those pages on which $x$ or $y$ occurs, and $f(x) = f(y) = \sqrt{N}$. Then $f(x) = f(y) = f(x,z) = f(y,z) = \sqrt{N}$, $f(z) = 2\sqrt{N}$, and $f(x,y) = 0$. This yields $e_G(x,y) = \infty$ and $e_G(x,z) = e_G(z,y) = 2/\log N$, which violates the triangle inequality for all $N$. It follows that the NWD is not a metric. Indeed, we should view the distance $e_G$ between two concepts as a relative similarity measure between those concepts. Then, while concept $x$ is semantically close to concept $y$ and concept $y$ is semantically close to concept $z$, concept $x$ can be semantically very different from concept $z$.

Another important property of the NWD is its *scale-invariance*. This means that, if the number $N$ of pages indexed by the search engine grows sufficiently large, the number of pages containing a given search term goes to a fixed fraction of $N$, and so does the number of pages containing conjunctions of search terms. This means that if $N$ doubles, then so do the $f$-frequencies. For the NWD to give us an objective semantic relation between search terms, it needs to become stable when the number $N$ of indexed pages grows. Some evidence that this actually happens was given in Example 3.4.

The NWD can be used as a tool to investigate the meaning of terms and the relations between them as given by common sense. This approach can be compared with the *Cyc* project [16], which tries to create artificial common sense. Cyc's knowledge base consists of hundreds of microtheories and hundreds of thousands of terms, as well as over a million hand-crafted assertions written in a formal language called CycL [21]. CycL is an enhanced variety of first order predicate logic. This knowledge base was created over the course of decades by paid human experts. It is therefore of extremely high quality. The Web, on the other hand, is almost completely unstructured, and offers only a primitive query capability that is not nearly flexible enough to represent formal deduction. But what it lacks in expressiveness the Web makes up for in size; Web search engines have already indexed more than ten billion pages and show no signs of slowing down. Therefore search engine databases represent the largest publicly-available single corpus of aggregate statistical and indexing information so far created, and it seems that even rudimentary analysis thereof yields a variety of intriguing possibilities. It is unlikely, however, that this approach can ever achieve 100% accuracy like in principle deductive logic can, because the Web mirrors humankind's own imperfect and varied nature. But, as we will see below, in practical terms the NWD can offer an easy way to provide results that are good enough for many applications, and which would be far too much work if not impossible to program in a deductive way.

In the following sections we present a number of applications of the NWD: hierarchical clustering and classification of concepts and names in a variety of domains, finding corresponding words in different languages, and a system that answers natural language questions.

### *3.4.2 Hierarchical Clustering*

To perform the experiments in this section, we used the *CompLearn* software tool [8], the same tool that has been used in Sect. 3.3 to construct trees representing hierarchical clusters of objects in an unsupervised way. However, now we use the normalized Web distance (NWD) instead of the normalized compression distance (NCD). Recapitulating, the method works by first calculating a distance matrix using NWD among all pairs of terms in the input list. Then it calculates a best-matching unrooted ternary tree using a novel quartet-method style heuristic based on randomized hill-climbing using a new fitness objective function optimizing the summed costs of all quartet topologies embedded in candidate trees [9].

#### 3.4.2.1 Colors and Numbers

In the first example [11], the objects to be clustered are search terms consisting of the names of colors, numbers, and some tricky words. The program automatically organized the colors towards one side of the tree and the numbers towards the other, Fig. 3.5. It arranges the terms which have as only meaning a color or a number, and nothing else, on the farthest reach of the color side and the number side, respectively. It puts the more general terms black and white, and zero, one, and two, towards the center, thus indicating their more ambiguous interpretation. Also, things which were not exactly colors or numbers are also put towards the center, like the word "small." We may consider this an example of automatic ontology creation.

#### 3.4.2.2 Dutch 17th Century Painters

In the example of Fig. 3.6, the names of fifteen paintings by Steen, Rembrandt, and Bol were entered [11]. The names of the associated painters were not included in the input, however they were added to the tree display afterwards to demonstrate the separation according to painters. This type of problem has attracted a great deal of attention [22]. A more classical solution would use a domain-specific database for similar ends. The present automatic oblivious method obtains results that compare favorably with the latter feature-driven method.

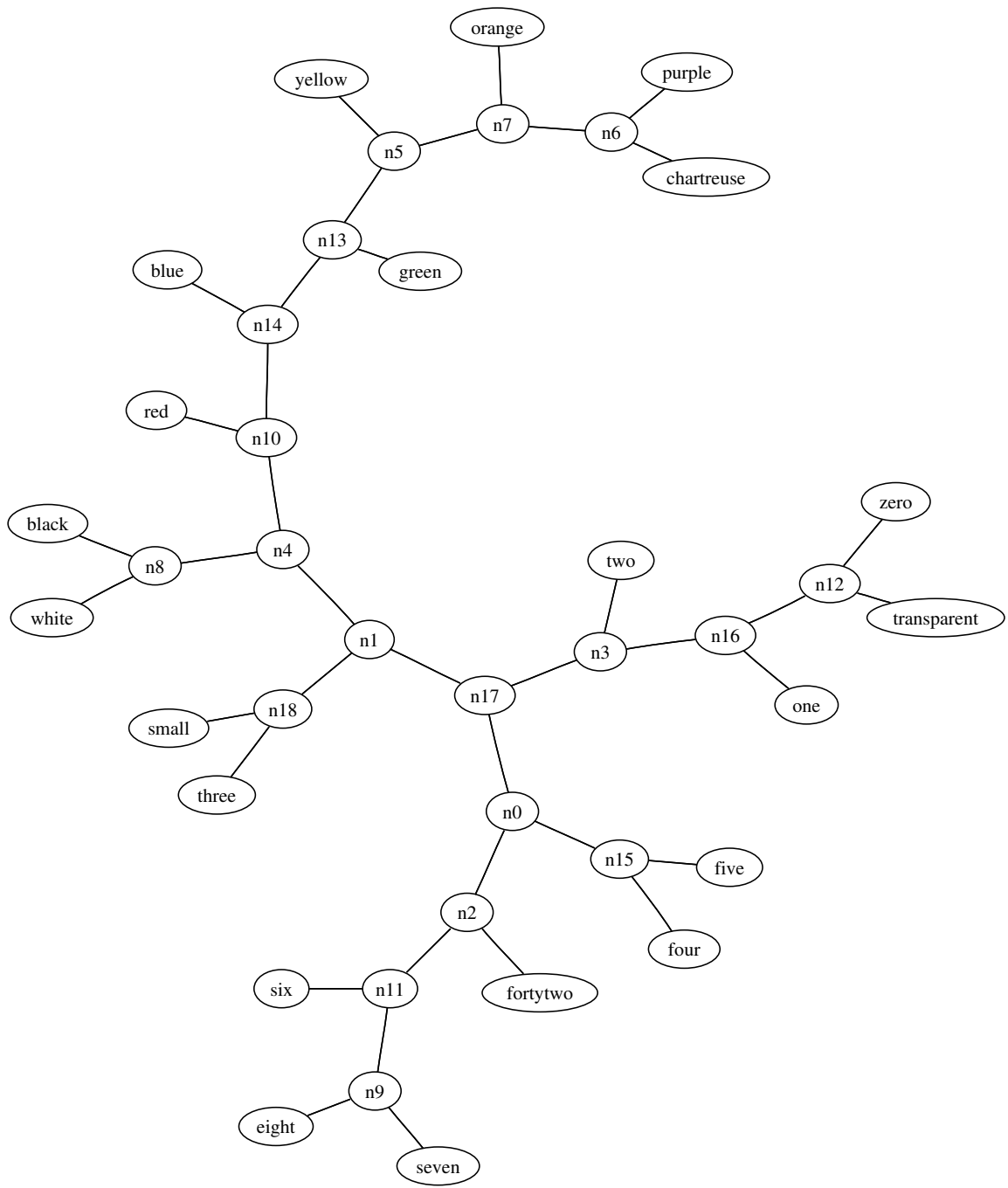

**Fig. 3.5** Colors, numbers, and other terms arranged into a tree based on the normalized Web distances between the terms

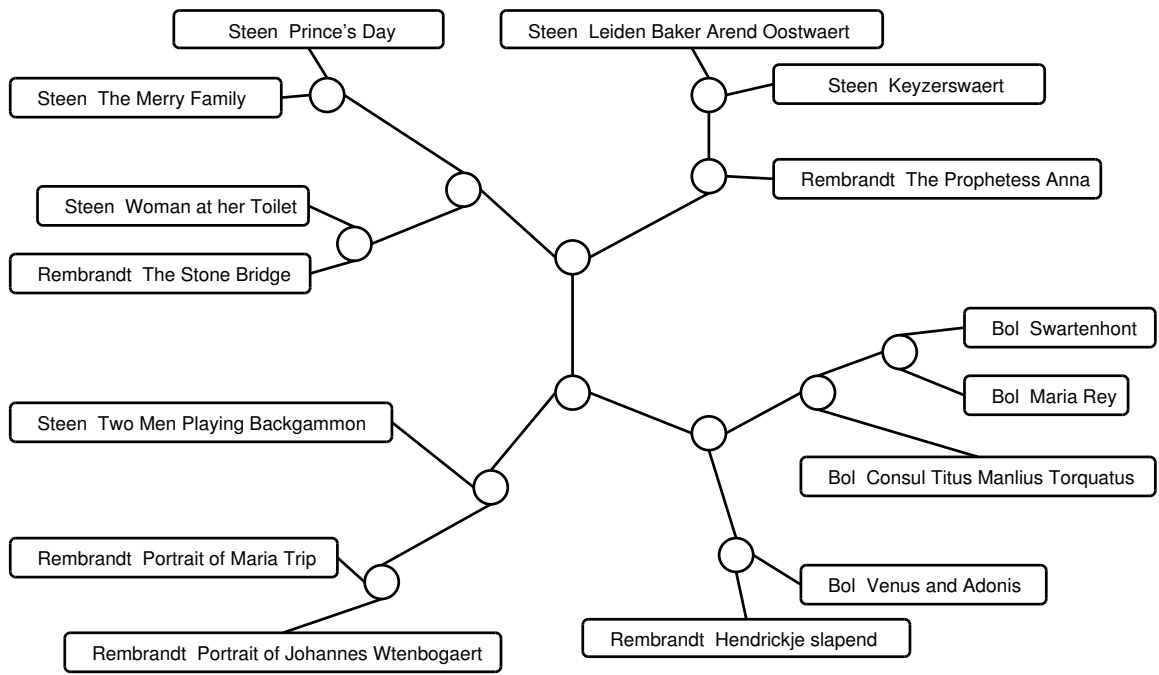

**Fig. 3.6** Fifteen paintings by three painters arranged into a tree by hierarchical clustering. To determine the normalized Web distances between the paintings, only the title names were used; the painter prefixes shown in the diagram were added afterwards to assist in interpretation

#### 3.4.2.3 Chinese Names

In the example of Fig. 3.7, several Chinese names were entered. The tree shows the separation according to concepts like regions, political parties, people, etc. See Fig. 3.8 for English translations of these names. The dotted lines with numbers inbetween each adjacent node along the perimeter of the tree represent the NWD values between adjacent nodes in the final ordered tree. The tree is presented in such a way that the sum of these values in the entire ring is minimized. This generally results in trees that makes the most sense upon initial visual inspection, converting an unordered binary tree to an ordered one. This feature allows for a quick visual inspection around the edges to determine the major groupings and divisions among coarse structured problems. It grew out of an idea originally suggested by Rutledge [22].

### 3.4.3 Support Vector Machine Learning

We augment the NWD method by adding a trainable component of the learning system. This allows us to consider classification rather than clustering problems. Here we use the Support Vector Machine (SVM) as a trainable component. For all SVM experiments, the LIBSVM software [5] has been used.

The setting is a binary classification problem on examples represented by search terms. We require a human expert to provide a list of at least 40 *training words*, consisting of at least 20 positive examples and

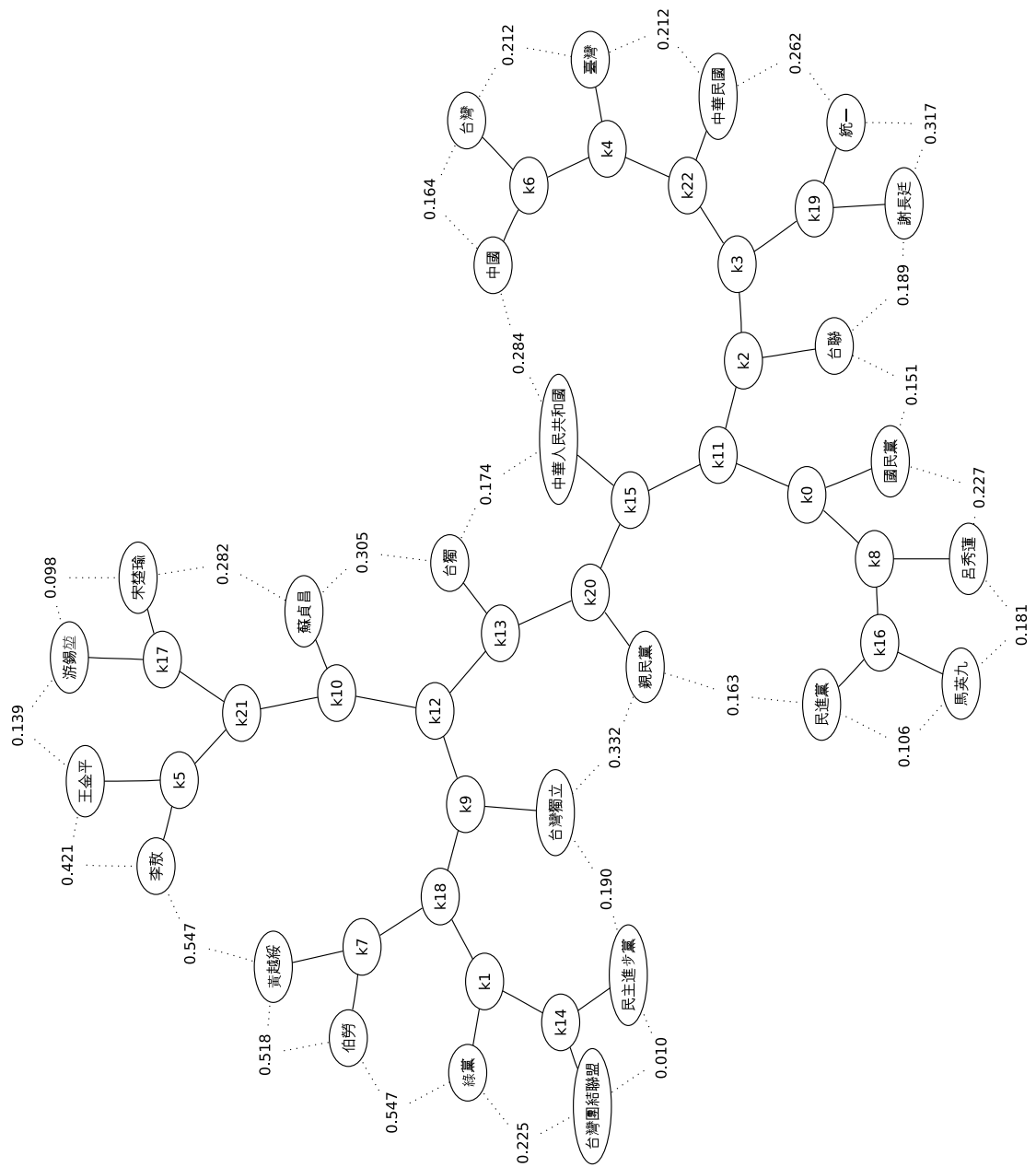

**Fig. 3.7** Names of several Chinese people, political parties, regions, and others. The nodes and *solid lines* constitute a tree constructed by a hierarchical clustering method based on the normalized Web distances between all names. The numbers at the perimeter of the tree represent NWD values between the nodes pointed to by the *dotted lines*. For an explanation of the names, refer to Fig. 3.8

中國 China
中華人民共和國 People's Republic of China
中華民國 Republic of China
伯勞 shrike (bird) [outgroup]
台灣 Taiwan (with simplified character "tai")
台灣團結聯盟 Taiwan Solidarity Union [Taiwanese political party]
台灣獨立 Taiwan independence
台獨 (abbreviation of the above)
台聯 (abbreviation of Taiwan Solidarity Union)
呂秀蓮 Annette Lu
國民黨 Kuomintang
宋楚瑜 James Soong
李敖 Li Ao
民主進步黨 Democratic Progressive Party
民進黨 (abbreviation of the above)
游錫堃 Yu Shyi-kun
王金平 Wang Jin-pyng
統一 unification [Chinese unification]
綠黨 Green Party
臺灣 Taiwan (with traditional character "tai")
蘇貞昌 Su Tseng-chang
親民黨 People First Party [political party in Taiwan]
謝長廷 Frank Hsieh
馬英九 Ma Ying-jeou
黃越綏 a presidential advisor and 2008 presidential hopeful

**Fig. 3.8** Explanations of the Chinese names used in the experiment that produced Fig. 3.7

20 negative examples, to illustrate the contemplated concept class. The expert also provides, say, six *anchor words* $a_1, \ldots, a_6$, of which half are in some way related to the concept under consideration. Then, we use the anchor words to convert each of the 40 training words $w_1, \ldots, w_{40}$ to 6-dimensional *training vectors* $\mathbf{v}_1, \ldots, \mathbf{v}_{40}$. The entry $v_{j,i}$ of $\mathbf{v}_j = (v_{j,1}, \ldots, v_{j,6})$ is defined as $v_{j,i} = e_G(w_i, a_j)$ ($1 \leq i \leq 40$, $1 \leq j \leq 6$). The training vectors are then used to train an SVM to learn the concept, and then test words may be classified using the same anchors and trained SVM model. We present all positive examples as *x*-data (input data), paired with $y = 1$. We present all negative examples as *x*-data, paired with $y = -1$.

The above method for transforming concepts into real valued vectors is not limited to be used with SVMs, but can be combined with any machine learning algorithm that can handle numeric inputs.

#### 3.4.3.1 Learning Prime Numbers

**Training Data**

**Positive examples** (21 cases)

| 11 | 13 | 17 | 19 | 2 |
| 23 | 29 | 3  | 31 | 37 |
| 41 | 43 | 47 | 5  | 53 |
| 59 | 61 | 67 | 7  | 71 |
| 73 |    |    |    |   |

**Negative examples** (22 cases)

| 10 | 12 | 14 | 15 | 16 |
| 18 | 20 | 21 | 22 | 24 |
| 25 | 26 | 27 | 28 | 30 |
| 32 | 33 | 34 | 4  | 6  |
| 8  | 9  |    |    |    |

**Anchors** (5 dimensions)
composite, number, orange, prime, record

**Testing Results**

|  | Positive tests | Negative tests |
|---|---|---|
| **Positive Predictions** | 101, 103, 107, 109, 79, 83, 89, 91, 97 | 110 |
| **Negative Predictions** |  | 36, 38, 40, 42, 44, 45, 46, 48, 49 |

**Accuracy:** 18/19 = 94.74%

**Fig. 3.9** NWD-SVM learning of prime numbers. All examples, i. e.,numbers, were converted into vectors containing the NWD values between that number and a fixed set of anchor concepts. The classification was then carried out on these vectors using a support vector machine. The only error made is classifying 110 as a prime

In Fig. 3.9 the method learns to distinguish prime numbers from non-prime numbers by example [11]. This example illustrates several common features the NWD method that distinguish it from the strictly deductive techniques. It is common for the classifications to be good but imperfect, and this is due to the unpredictability and uncontrolled nature of the Web distribution.

#### 3.4.3.2 WordNet Semantics: Learning Religious Terms

The next example (see the preliminary version of [11]) has been created using WordNet [12], which is a semantic concordance of English. It also attempts to focus on the meaning of words instead of the word itself. The category we want to learn here is termed "religious" and represents anything that may pertain to religion. The negative examples are constituted by simply everything else (see Fig. 3.10). Negative examples were chosen randomly and uniformly from a dictionary of English words. This category represents a typical expansion of a node in the WordNet hierarchy. The accuracy on the test set is 88.89%.

#### 3.4.3.3 WordNet Semantics: One Hundred Experiments

The previous example shows only one hand-crafted special case. To investigate the more general statistics, a method was devised to estimate how well the NWD-SVM approach agrees with WordNet in a large number of automatically selected semantic categories [11].

Before we explain how each category was automatically generated, we first review the structure of Word-Net; the following is paraphrased from the official WordNet documentation available online. WordNet is

**Training Data**

**Positive examples** (22 cases)

| | | | | |
|---|---|---|---|---|
| Allah | Catholic | Christian | Dalai Lama | God |
| Jerry Falwell | Jesus | John the Baptist | Mother Theresa | Muhammad |
| Saint Jude | The Pope | Zeus | bible | church |
| crucifix | devout | holy | prayer | rabbi |
| religion | sacred | | | |

**Negative examples** (23 cases)

| | | | | |
|---|---|---|---|---|
| Abraham Lincoln | Ben Franklin | Bill Clinton | Einstein | George Washington |
| Jimmy Carter | John Kennedy | Michael Moore | atheist | dictionary |
| encyclopedia | evolution | helmet | internet | materialistic |
| minus | money | mouse | science | secular |
| seven | telephone | walking | | |

**Anchors** (6 dimensions)

| | | | | | |
|---|---|---|---|---|---|
| evil | follower | history | rational | scripture | spirit |

**Testing Results**

| | **Positive tests** | **Negative tests** |
|---|---|---|
| **Positive Predictions** | altar, blessing, communion, heaven, sacrament, testament, vatican | earth, shepherd |
| **Negative Predictions** | angel | Aristotle, Bertrand Russell, Greenspan, John, Newton, Nietzsche, Plato, Socrates, air, bicycle, car, fire, five, man, monitor, water, whistle |

**Accuracy:** 24/27 = 88.89%

**Fig. 3.10** NWD-SVM learning of religious terms. All training and test examples were converted into vectors containing the NWD values between that example concept and a fixed set of anchor concepts. The classification was then carried out on these vectors using a support vector machine

called a semantic concordance of the English language. It seeks to classify words into many categories and interrelate the meanings of those words. WordNet contains synsets. A synset is a synonym set; a set of words that are interchangeable in some context, because they share a commonly-agreed upon meaning with little or no variation. Each word in English may have many different senses in which it may be interpreted; each of these distinct senses points to a different synset. Every word in WordNet has a pointer to at least one synset. Each synset, in turn, must point to at least one word. Thus, we have a many-to-many mapping between English words and synsets at the lowest level of WordNet. It is useful to think of synsets as nodes in a graph. At the next level we have lexical and semantic pointers. Lexical pointers are not investigated in this section; only the following semantic pointer types are used in our comparison: A semantic pointer is simply a directed edge in the graph whose nodes are synsets. The pointer has one end we call a *source* and the other end we call a *destination*. The following relations are used:

1. *hyponym*: $X$ is a hyponym of $Y$ if $X$ is a (kind of) $Y$.
2. *part meronym*: $X$ is a part meronym of $Y$ if $X$ is a part of $Y$.
3. *member meronym*: $X$ is a member meronym of $Y$ if $X$ is a member of $Y$.
4. *attribute*: A noun synset for which adjectives express values. The noun *weight* is an attribute, for which the adjectives *light* and *heavy* express values.

5. *similar to*: A synset is similar to another one if the two synsets have meanings that are substantially similar to each other.

Using these semantic pointers we may extract simple categories for testing. First, a random semantic pointer (or edge) of one of the types above is chosen from the WordNet database. Next, the source synset node of this pointer is used as a root. Finally, we traverse outward in a breadth first order starting at this root and following only edges that have an identical semantic pointer type; that is, if the original semantic pointer was a hyponym, then we would only follow hyponym pointers in constructing the category. Thus, if we were to pick a hyponym link initially that says a *tiger* is a *cat*, we may then continue to follow further hyponym relationships in order to continue to get more specific types of cats. See the WordNet homepage [20] documentation for specific definitions of these technical terms.

Once a category is determined, it is expanded in a breadth first way until at least 38 synsets are within the category. 38 was chosen to allow a reasonable amount of training data to be presented with several anchor dimensions, yet also avoiding too many. Here, a rule of thumb is helpful: it states that the number of dimensions in the input data must not exceed one tenth the number of training samples. If the category cannot be expanded this far, then a new one is chosen. Once a suitable category is found, and a set of at least 38 members has been formed, a training set is created using 25 of these cases, randomly chosen. Next, three are chosen randomly as anchors. And finally the remaining ten are saved as positive test cases. To fill in the negative training cases, random words are chosen from the WordNet database. Next, three random words are chosen as unrelated anchors. Finally, 10 random words are chosen as negative test cases.

For each case, the SVM is trained on the training samples, converted to 6-dimensional vectors using NWD. The SVM is trained on a total of 50 samples. The kernel-width and error-cost parameters are automatically determined using five-fold cross validation. Finally testing is performed using 20 examples in a balanced ensemble to yield a final accuracy.

There are several caveats with this analysis. It is necessarily rough, because the problem domain is difficult to define. There is no protection against certain randomly chosen negative words being accidentally members of the category in question, either explicitly in the greater depth transitive closure of the category, or perhaps implicitly in common usage but not indicated in WordNet. Another detail to notice is that WordNet is available through some Web pages, and so undoubtedly contributes something to Web page counts. Further experiments comparing the results when filtering out WordNet images on the Web suggest that this problem does not usually affect the results obtained, except when one of the anchor terms happens to be very rare and thus receives a non-negligible contribution towards its page count from WordNet views. In general, the previous NCD based methods, as in [9], exhibit large-granularity artifacts at the low end of the scale; for small strings we see coarse jumps in the distribution of NCD for different inputs which makes differentiation difficult. With the Web based NWD we see similar problems when page counts are less than a hundred.

We ran 100 experiments. The histogram of agreement accuracies is shown in Fig. 3.11. On average, the NWD method turns out to agree well with the WordNet semantic concordance made by human experts. The mean of the accuracies of agreements is 0.8725. The variance is approximately 0.01367, which gives a standard deviation of approximately 0.1169. Thus, it is rare to find agreement less than 75%.

We conclude this section with a more abstract view of the NWD-SVM method. As we have seen, this method does not use an individual word in isolation, but instead uses an ordered list of its NWD relationships with fixed anchors. Therefore nothing can be attached to the isolated interpretation of a literal term, but only to the ordered list by which it is represented. That is to say, the inputs to our SVM are not directly search terms, but instead an image of the search term through the lens of the Web distribution, and relative to other fixed terms which serve as a grounding for the term. In most schools of ontological thought, and indeed in the WordNet database, there is imagined a two-level structure that characterizes language: a many-to-many relationship between word-forms or utterances and their many possible meanings. Each link in this association will be represented in the Web distribution with strength proportional to how common that usage is found on the Web. The NWD then amplifies and separates the many contributions towards the aggregate page count sum, thereby revealing some components of the latent semantic Web. In almost every informal theory of cognition we have the idea of connectedness of different concepts in a network, and this is precisely the structure that the NWD experiments attempt to explore.

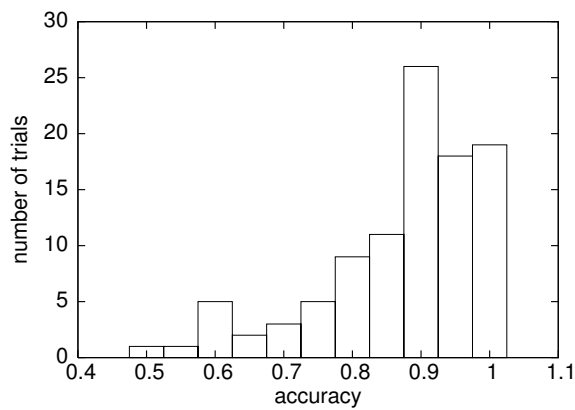

**Fig. 3.11** Histogram of accuracies over 100 trials of WordNet experiment. The average accuracy achieved in the experiments is 0.8725

### 3.4.4 Matching the Meaning

Yet another potential application of the NWD method is in natural language translation. In the experiment below [11] we do not use SVMs to obtain the result, but determine correlations instead. Suppose we are given a system that tries to infer a translation-vocabulary among English and Spanish. Assume that the system has already determined that there are five words that appear in two different matched sentences, but the permutation associating the English and Spanish words is, as yet, undetermined. This setting can arise in real situations, because English and Spanish have different rules for word-ordering. Thus, at the outset we assume a pre-existing vocabulary of eight English words with their matched Spanish translation. Can we infer the correct permutation mapping the unknown words using the pre-existing vocabulary as a basis?

We start by forming an NWD matrix using additional English words of which the translation is known, Fig. 3.12. We label the columns by the translation-known English words, the rows by the translation-unknown words. The entries of the matrix are the NWDs of the English words labeling the columns and rows. This constitutes the English basis matrix. Next, consider the known Spanish words corresponding to the known English words. Form a new matrix with the known Spanish words labeling the columns in the same order as the known English words. Label the rows of the new matrix by choosing one of the many possible permutations of the unknown Spanish words. For each permutation, form the NWD matrix for the Spanish words, and compute the pairwise correlation of this sequence of values to each of the values in the given English word basis matrix. Choose the permutation with the highest positive correlation. If there is no positive correlation, report a failure to extend the vocabulary. In this example, the computer inferred the correct permutation for the testing words, see the right table in Fig. 3.12.

466Paul M. B. Vitányi, Frank J. Balbach, Rudi L. Cilibrasi, and Ming Li

| Given starting vocabulary | | Unknown permutation | | Predicted permutation | |
|---|---|---|---|---|---|
| **English** | **Spanish** | **English** | **Spanish** | **English** | **Spanish** |
| tooth | diente | plant | bailar | plant | planta |
| joy | alegria | car | hablar | car | coche |
| tree | arbol | dance | amigo | dance | bailar |
| electricity | electricidad | speak | coche | speak | hablar |
| table | tabla | friend | planta | friend | amigo |
| money | dinero | | | | |
| sound | sonido | | | | |
| music | musica | | | | |

**Fig. 3.12** English-Spanish translation problem. The left table is given to the system as background knowledge, the middle table contains words whose mapping is not known. The right table shows the mapping determined by the system

### 3.4.5 Question-Answer System

A typical procedure for finding an answer on the World Wide Web consists in entering some terms regarding the question into a Web search engine and then browsing the search results in search for the answer. This is particularly inconvenient when one uses a mobile device with a slow internet connection and small display. Question-answer (QA) systems attempt to solve this problem. They allow the user to enter a question in natural language and generate an answer by searching the Web autonomously.

In this section we describe some parts of the QA system QUANTA [25] that uses a variant of the NID to identify the correct answer to a question out of several candidates for answers. QUANTA is remarkable in that it uses neither NCD nor NWD introduced so far, but a variation that is nevertheless based on the same theoretical principles. This variation is tuned to the particular needs of a QA system. We begin with some motivation for this particular variant and then describe it formally within the Kolmogorov framework. We shall focus on the new distance measure, not on other, interesting, details of the system, such as parsing and chunking the question, interfacing Web search engines, or generating candidate answers. We thus assume that for a given question a set of possible answers is already available, and the system only has to pick the (or a) right one.

As an example we consider the question "Which city is Lake Washington by?," which allows for many answers, among them Seattle, Bellevue, or Dallas. The first two cities are correct answers, but the preferred answer would be Seattle as the more well-known city. In a straightforward attempt to finding the right answer using the normalized Web distance we could compute $e_G$(Lake Washington, Bellevue), $e_G$(Lake Washington, Seattle) and $e_G$(Lake Washington, Dallas) and pick the city with the least distance. An experiment performed in February 2008 with a popular Web search engine yielded

- $e_G$(Lake Washington, Bellevue) $= 0.4658$,
- $e_G$(Lake Washington, Seattle) $= 0.5716$,
- $e_G$(Lake Washington, Dallas) $= 0.8302$,

so that Bellevue would have been chosen. Without normalization the respective distance values are 6.33, 7.54 and 10.95. Intuitively, the reason for Seattle being relatively far away from Lake Washington (in terms of $e_G$) is that, due to Seattle's size and popularity, it has many concepts in its neighborhood not all of which can be close. For the less known city of Bellevue, however, Lake Washington is relatively more important. Put differently, the concept "Seattle" contains a lot of information that is irrelevant for its being situated at Lake Washington. Symmetrically, Lake Washington encompasses much information unrelated to Seattle. A variation of (3.1) that accounts for possible irrelevant information is then

3 Normalized Information Distance

$$E_{\min,0}(x,y) = \min\{\ell(p) : U(x,p,r) = y \text{ and } U(y,p,q) = x \quad (3.7)$$
$$\text{and } \ell(p) + \ell(q) + \ell(r) \leq E_0(x,y)\}.$$

Here, $r$ represents the irrelevant information in $y$ and $q$ the irrelevant information in $x$. The additional restriction $\ell(p) + \ell(q) + \ell(r) \leq E_0(x,y)$ ensures that the amount of irrelevant information is limited. Without it, one could set $r = y$ and $q = x$ and always use a program $p$ of constant size that merely outputs one of its arguments.

Similarly as $E_0$ in (3.1), $E_{\min,0}$ cannot be used practically right away, it must be converted into a formula based on $K$ [25]:

**Theorem 3.9.** $E_{\min,0}(x,y) = \min\{K(x|y), K(y|x)\} + O(\log \ell(xy))$.

The term $\min\{K(x|y), K(y|x)\}$ is also called the *min distance* and denoted by $E_{\min}$. The min distance is not a metric since it does not satisfy the triangle inequality. But in question-answer systems on the internet, distances are measured with partial information anyway, hence it is unreasonable to require the triangle inequality to hold. Furthermore, $E_{\min}$ satisfies the density conditions in (3.2) only for strings $x$ with $K(x) \geq \ell(x) + O(1)$. It does not hold for objects with a low Kolmogorov complexity, which correspond to concepts with high frequency, such as Seattle. That $E_{\min}$ violates (3.2) for such objects intuitively means that popular concepts are allowed to have a denser neighborhood. This property is therefore rather a feature of $E_{\min}$ than a bug.

In another step paralleling the development of the NID, the min distance can be normalized. Analogously to $e$, we define the normalized version $e_{\min}$ of $E_{\min}$ as

$$e_{\min}(x,y) = \frac{E_{\min}(x,y)}{\min\{K(x), K(y)\}} = \frac{\min\{K(x|y), K(y|x)\}}{\min\{K(x), K(y)\}}.$$

It follows, though not obviously, that $e_{\min}(x,y) \leq e(x,y)$ for all $x$ and $y$.

The normalized min distance $e_{\min}$ can be approximated by Web statistics in the same way as the NWD approximates the NID (cf. (3.6)), namely using the formula

$$e_{G,\min}(x,y) = \frac{\min\{\log f(x), \log f(y)\} - \log f(x,y)}{\log N - \max\{\log f(x), \log f(y)\}}.$$

Applying this normalized min Web distance to our above example question and answers, we obtain:

- $e_{G,\min}(\text{Lake Washington, Bellevue}) = 0.4496$,
- $e_{G,\min}(\text{Lake Washington, Seattle}) = 0.4281$,
- $e_{G,\min}(\text{Lake Washington, Dallas}) = 0.7746$,

that is, the answer "Seattle" would now be preferred over "Bellevue," and Dallas is still out of the question.

Regardless of whether we used $e_G$ or $e_{G,\min}$, statistics would be obtained for the (co-)occurrence of the following single words and pairs:

- "Lake Washington,"
- "Seattle,"
- "Bellevue,"
- "Lake Washington" and "Seattle,"
- "Lake Washington" and "Bellevue."

But there is nothing hinting to the fact that we are looking for the co-location of a city and a lake. Of course, in this example it is reasonably clear. If, however, the question was represented by "Alan Turing," and candidate answers were "London," "Wilmslow," and "Paris," it would be unclear whether we are looking for Turing's place of birth, place of death, or any other place related to him. Clearly, the veracity of any of these answers depends on the particular question. It is therefore necessary to add some clues as to what the question is to the queries given to the Web search engine. For example adding the phrase "is born in" to all queries would (hopefully) limit the obtained statistics to Web pages that are concerned with Turing's birth and therefore result in "London" being chosen as answer.

The idea of such side information can easily be incorporated into the underlying theory by adding a condition to all terms in $e_{\min}$ (or $e$ for that matter), yielding

$$e_{\min}(x,y|c) = \frac{\min\{K(x|y,c), K(y|x,c)\}}{\min\{K(x|c), K(y|c)\}},$$

where $c$ denotes the conditional information, such as "is born in" in the above example. The extraction of a suitable $c$ requires some sophistication and is beyond the scope of our discussion here.

The conditional version of $e_{\min}$ is at the core of the QUANTA system, whose question answering capabilities compare favorably with other QA systems [25]. The beneficial properties of $e_{\min}$ can perhaps best seen in comparison to other measures such as the normalized max distance $e$ or the unnormalized distances $E$ and $E_{\min}$. Replacing $e_{\min}$ with $e$ results in answers that are still technically correct but often less popular and therefore less "good." We already mentioned Bellevue being preferred over Seattle as a city located at Lake Washington. Another example is the question "When was CERN founded?," which would be answered by $e$ with "52 years ago," correct in 2006, whereas $e_{\min}$ responds more accurately with "1954."

Using the unnormalized $E$ gives overly much weight to popular concepts. For instance, "Who is the greatest scientist of all?" would be answered with "God," whereas $e_{\min}$ would give "Newton," the reason for this discrepancy being that, in terms of Web pages, God is much more popular than Newton. More generally, experiments have shown [25] that $E_{\min}$ and $E$ perform about 8% worse than $e_{\min}$.

The development of $e_{\min}$ to pick the most plausible answer in a QA system demonstrates how distance measures customized to special applications can be derived from first principles of Kolmogorov complexity theory, which in turn shows the power and flexibility of this theoretical approach.

### 3.5 Conclusions

The approach described in this chapter rests upon the simple idea that an *information distance* between two objects can be measured by the size of the shortest description that transforms each object into the other one. This idea is most naturally expressed mathematically using Kolmogorov complexity. Kolmogorov complexity, moreover, provides mathematical tools to show that such a measure is, in a proper sense, universal among all (upper semi)computable distance measures satisfying a natural density condition. These comprise most, if not all, distances one may be interested in. This information distance happens to be a metric. Since two large, very similar, objects may have the same information distance as two small, very dissimilar, objects, in terms of similarity it is the relative distance we are interested in. Hence we normalize the information metric a relative similarity (also metric) in between 0 and 1. However, the normalized information metric is uncomputable. We approximate its Kolmogorov complexity parts by off-the-shelve compression programs (in the case of the normalized compression distance) or readily available statistics from the internet (in case of the normalized Web distance). The outcome are two practical distance measures, for literal as well as for non-literal data, that have been proved useful in numerous applications, some of which have been presented in the previous sections.

Just as important as the successes of these practical measures, however, is the underlying process used to derive them. The derivations of NCD and NWD are special instances of this process, which can roughly be broken into three steps: (1) devising an abstract distance notion, (2) transforming it inside the abstract mathematical realm into an equivalent, yet more easily realizable, formula, and (3) using real-world algorithms or data to practically realize the theoretically conceived measure. That this approach does not work by chance just for the information distance, is demonstrated by the derivation of the minimum distance, which employs the same three step process, just with different starting requirements for the distance measure.

Central design principles behind these Kolmogorov-based distance measures are the requirement of universality and the use of absolute measures of information content to achieve this universality. From these principles it follows naturally that the resulting distance measures are independent of fixed feature sets and do not require parameters for tuning. They can thus be used to build feature- and parameter-free methods that are suited for many tasks in exploratory data mining, alignment-free genomics, and elsewhere.

## Appendix

### List of Symbols

- $\bar{x}$: self-delimiting encoding of string $x$
- $\ell(\cdot)$: length of a string
- $\varepsilon$: empty string
- $|\ldots|$: cardinality
- $\mathscr{R}^+$: set of all non-negative rational numbers
- $\mathscr{R}$: set of all rational numbers
- $\mathscr{N}$: set of all natural numbers
- $\mathscr{Q}$: set of all rational numbers
- $\lfloor\ldots\rfloor$: floor function
- $\langle\ldots\rangle$: pairing function
- $C(\cdot|\cdot)$: conditional Kolmogorov complexity
- $C(\cdot)$: unconditional Kolmogorov complexity
- $K(\cdot|\cdot)$: conditional Kolmogorov prefix complexity
- $K(\cdot)$: unconditional Kolmogorov prefix complexity
- **m**: universal upper semicomputable discrete semimeasure
- $\oplus$: bitwise xor
- $E$: max distance
- $E_0$: information distance
- $e$: normalized information distance
- $e_Z$: normalized compression distance
- $e_G$: normalized Web distance
- **W**: set of all Web pages
- **x**: set of all Web pages containing term $x$
- $f(x)$: number of Web pages containing term $x$
- $g(x)$: probability that a Web page contains term $x$

- $E_{\min}$: min distance
- $e_{\min}$: normalized min distance
- $e_{G,\min}$: normalized min Web distance